# A New Class of Nonsymmetric Multivariate Dependence Measures

*Hui Li*[1]

Following our previous work on copula-based nonsymmetric bivariate dependence measures, we propose a new set of conditions on nonsymmetric multivariate dependence measures which characterize both independence and complete dependence of one random variable on a group of random variables. The measures are nonparametric in that they are copula-based and are invariant under continuous bijective transformations on the group of random variables. We also construct explicitly new measures that satisfy the conditions. Besides, we extend the $*$ product on bivariate copulas to multivariate copulas and prove the DPI condition and self-equitability for the new measures. A further extension to measures of dependence of one group of random variables on another group of random variables is also discussed.

## 1. Introduction

In a previous paper (Li, 2015), we proposed a new set of conditions on copula-based nonsymmetric bivariate dependence measures such that a measure captures both independence and complete dependence of one random variable on another random variable, which is also invariant under continuous bijective transformations on the other random variable. As complete dependence is not symmetric in a functional relationship, the new measures are also nonsymmetric, which may be applied to, for example, regression dependence order as shown by Dette, Siburg and Stoimanov (2010). As regression dependence is not restricted to just bivariate random variables, we will extend our work to characterize measures of dependence of one random variable on a group of random variables. More specifically, the new type of measure should be again nonsymmetric, takes value zero if and only if the one random variable is independent of the group of random variables even if the random variables within the group are correlated with each other, takes value one if and only if the one random variable is a function of the group of random variables, and is invariant under continuous bijective transformations on the group of random variables. To our knowledge, this kind of measure has not been introduced before (see, for example, the review article by Schmid, et al 2010) and it could be very useful in the analysis of large amount of data and complex relationships. The last property where the measure is invariant under transformations of the group of random variables is very helpful as the complex relationships may not be measured in the right units and may not be easily parameterized. We note that Sanchez and Trutschnig (2014) discussed a multidimensional extension to some nonsymmetric bivariate dependence measures. But their extension is still bivariate in nature as it averages over the bivariate dependence measure for each of the group of random variables on the one random variable, and complete dependence in their extension means each of the group of random variables is completely dependent on the one random variable. Instead, our measures determine the dependence relationship of one random variable on the group. Again we will explore copula-based dependence measures as copula contains the dependence information of continuous random variables.

[1]Email: lichtung2014@gmail.com. The views expressed in this article are the author's own and do not represent the views of his employer.

The structure of the current paper is as follows. Section 2 reviews the original axioms of the symmetric dependence measure proposed by Rényi (1959) and those for the nonsymmetric dependence measure proposed in our previous paper (Li, 2015). We then present our new axioms for dependence measure of one random variable on a group of random variables. Section 3 reviews the basic properties of multivariate copula which forms the basis for the study of the new measures. Section 4 presents a new class of nonsymmetric multivariate dependence measure and shows that it satisfies the new axioms. Section 5 discusses the Data Processing Inequality (DPI) through the generalized $*$ product on copulas, which is used to prove invariance under bijective transformations or the self-equitable property of the dependence measures. Section 6 discusses a further extension to measures of dependence of one group of random variables on another group of random variables. Section 7 concludes the paper.

## 2. Axioms of bivariate and multivariate dependence measures

In 1959, Rényi (1959) introduced a set of axioms as the criteria of a symmetric nonparametric measure of dependence $R(X,Y)$ for two random variables $X,Y$ on a common probability space.

a) $R(X,Y)$ is defined for all non-constant random variables $X,Y$;
b) $R(X,Y)=R(Y,X)$;
c) $0\leq R(X,Y)\leq 1$;
d) $R(X,Y)=0$ if and only if $X,Y$ are independent;
e) $R(X,Y)=1$ if either $Y=f(X)$ or $X=g(Y)$ almost surely for some Borel-measurable functions $f,g$;
f) If $f$ and $g$ are Borel-measurable bijections on $\mathbb{R}$, then $R(f(X),g(Y))=R(X,Y)$;
g) If $X,Y$ are jointly normal with correlation coefficient ρ, then $R(X,Y)=|\rho|$.

It is well-known that although independence is a symmetric property, complete dependence is not. If $X$ is a function of $Y$, then $X$ is completely dependent on $Y$ but $Y$ need not to be completely dependent on $X$ unless the function is a bijection. So Rényi's condition b) is somewhat unintuitive. Recently, nonsymmetric measures of dependence have started to attract some attentions as new research on properties of copula naturally leads to them, see Dette, Siburg and Stoimenov (2010), Trutschnig (2011). A systematic study along the similar lines of Rényi's work was started in a previous paper (Li, 2015). Specifically, we assume $R(X,Y)$ measures the degree of dependence of $Y$ on $X$ and satisfies the following conditions:

a') $R(X,Y)$ is defined for all continuous random variables $X,Y$;

b') $R(X,Y)$ may not be equal to $R(Y,X)$;

c') $0 \leq R(X,Y) \leq 1$;

d') $R(X,Y) = 0$ if and only if $X,Y$ are independent;

e') $R(X,Y) = 1$ if and only if $Y = f(X)$ almost surely for a Borel-measurable function $f$;

f') If $g$ is a continuous bijection on $\mathbb{R}$, then $R(g(X),Y) = R(X,Y)$;

Condition a") restricts the random variables to continuous ones such that the copula between them is uniquely defined. Condition b") specifies that the dependence measure can be nonsymmetric. But if a copula is symmetric or $C(u,v) = C(v,u)$, then $R(X,Y) = R(Y,X)$. Conditions c") and d") are the same as Rényi's conditions c) and d) as independence is a symmetric property. Condition e") is more explicit about the nonsymmetric nature of dependence and is stronger as $R(X,Y) = 1$ happens if and only if $Y = f(X)$ almost surely. As a nonsymmetric measure, condition f'') only requires the measure to be invariant under bijective transformations on $X$. We may require $g$ to be continuous such that the copula of $g(X),Y$ is uniquely defined.

Now we will propose an extension from bivariate dependence measure to the multivariate case where a random variable may depend on a group of random variables. The extended conditions are as follows:

a") $R(X_1, X_2, \dots, X_n, Y)$ is defined for all continuous random variables $X_1, X_2, \dots, X_n, Y$;

b") $R(X_1, X_2, \dots, X_n, Y)$ may not be equal to $R(P(X_1, X_2, \dots, X_n, Y))$, where $P$ is a non-trivial permutation of $Y\ with\ X_1, X_2, \dots, X_n$;

c") $0 \leq R(X_1, X_2, \dots, X_n, Y) \leq 1$;

d") $R(X_1, X_2, \dots, X_n, Y) = 0$ if and only if $Y$ is independent of $X_1, X_2, \dots, X_n$;

e") $R(X_1, X_2, \dots, X_n, Y) = 1$ if and only if $Y = f(X_1, X_2, \dots, X_n)$ almost surely for a Borel-measurable function $f: \mathbb{R}^n \to \mathbb{R}$;

f") If $g$ is a continuous bijection on $\mathbb{R}^n$, then $R(g(X_1, X_2, \dots, X_n), Y) = R(X_1, X_2, \dots, X_n, Y)$;

$R(X_1, X_2, \dots, X_n, Y)$ measures the dependence of $Y$ on $X_1, X_2, \dots, X_n$. Condition a”) requires the random variables to be continuous such that the copula between them is uniquely defined. Condition b”) indicates that we are measuring the dependence of $Y$ on $X_1, X_2, \dots, X_n$ such that the measure is not symmetric. Condition d”) and condition e”) do not require that $X_1, X_2, \dots, X_n$ be independent of each other. Condition f”) require $g$ to be continuous such that the copula of $g(X_1, X_2, \dots, X_n), Y$ is uniquely defined. It also implies that permutation within $X_1, X_2, \dots, X_n$ does not change the dependence measure.

Axioms for symmetric multivariate dependence measures have been reviewed in Schmid, et al (2010). But they are not relevant here as symmetric measures may not be able to characterize complete dependence as discussed in Li (2015).

## 3. Basic properties of multivariate copula

Let $I$ denote the closed unit interval $[0,1]$ and $I^n$ the closed unit n-cube $[0,1] \times \cdots \times [0,1]$.

DEFINITION 3.1. A multivariate copula is a function $C: I^n \to I$ that satisfies the following conditions (Nelson, 2006):

(i) for $\boldsymbol{u} \in I^n, C(\boldsymbol{u}) = 0$ if at least one coordinate of $\boldsymbol{u}$ (bold-face marks a vector) is 0.

(ii) for $\boldsymbol{u} \in I^n, C(\boldsymbol{u}) = u_k$ if all coordinates of $\boldsymbol{u}$ equal to 1 except $u_k$.

(iii) for all $\boldsymbol{a}, \boldsymbol{b} \in I^n$ such that $a_i \le b_i$ for all i, $V_C([\boldsymbol{a}, \boldsymbol{b}]) \ge 0$, where the $C$-volume

$$V_C([\boldsymbol{a}, \boldsymbol{b}]) = \sum sgn(\boldsymbol{c}) C(\boldsymbol{c}) \tag{1}$$

sums over all vertices $\boldsymbol{c}$ of the n-box $B = [\boldsymbol{a}, \boldsymbol{b}]$ and

$$sgn(\boldsymbol{c}) = \begin{cases} 1, \; if \; c_k = a_k \; for \; an \; even \; number \; of \; k's, \\ -1, \; if \; c_k = a_k \; for \; an \; odd \; number \; of \; k's. \end{cases}$$

Equivalently, the $C$-volume of the n-box $B = [\boldsymbol{a}, \boldsymbol{b}]$ is the $n$th order difference of $C$ on $B$

$$V_C(B) = \Delta_{\boldsymbol{a}}^{\boldsymbol{b}} C(\boldsymbol{t}) = \Delta_{a_n}^{b_n} \Delta_{a_{n-1}}^{b_{n-1}} \cdots \Delta_{a_2}^{b_2} \Delta_{a_1}^{b_1} C(\boldsymbol{t}) \tag{2}$$

where the n first order differences are defined as

$$\Delta_{a_k}^{b_k} C(\boldsymbol{t}) = C(t_1, \cdots, t_{k-1}, b_k, t_{k+1}, \cdots, t_n) - C(t_1, \cdots, t_{k-1}, a_k, t_{k+1}, \cdots, t_n) \tag{3}$$

Note that the $C$-volume actually induces a *stochastic* measure $\mu_C$ on the Lebesgue $\sigma$–algebra for $I^n$ such that

$$\mu_C\left([0,1]^{i-1} \times A \times [0,1]^{n-i}\right) = \lambda(A) \tag{4}$$

where $\lambda$ is the Lebesgue measure on $\mathbb{R}$ and $A$ is a Lebesgue measurable set in [0,1]. We can also define the $k$th sub- $C$-volume on the k-box $B_k = [a_1, b_1] \times \cdots \times [a_k, b_k]$ as

$$V_C^k(B_k, t_{k+1}, \cdots, t_n) = \Delta_{a_k}^{b_k} \Delta_{a_{k-1}}^{b_{k-1}} \cdots \Delta_{a_2}^{b_2} \Delta_{a_1}^{b_1} C(t_1, \cdots, t_k, t_{k+1}, \cdots, t_n) \tag{5}$$

Note that $V_C^k(B_k, t_{k+1}, \cdots, t_n)$ is non-negative and is non-decreasing in $(t_{k+1}, \cdots, t_n)$. For example,

$$0 \leq V_C^k(B_k, x) \leq V_C^k(B_k, y) \quad \text{if } x \leq y \tag{6}$$

This can be proved as a copula is non-decreasing in each argument.

PROPOSITION 3.2. $|C(\boldsymbol{v}) - C(\boldsymbol{u})| \leq \sum_1^n |v_k - u_k|$, such that $C$ is uniformly continuous.

A copula is just a joint distribution function with uniform marginals $U_1, \cdots, U_n$. Copulas are of interest because they link general one-dimensional marginal distributions to joint distributions. Sklar (1959) showed that, for any real random variables $X_1, X_2, \dots, X_n$ with continuous marginal distribution functions $F_1, F_2, \dots, F_n$ and joint distribution function $F$, there is a unique copula $C$ such that

$$F(x_1, x_2, \dots, x_n) = C(F_1(x_1), F_2(x_2), \dots, F_n(x_n)) \tag{7}$$

Let $\mathbb{C}^n$ be the set of all n-variate copulas. Any copula $C \in \mathbb{C}^n$ can be decomposed into the sum of an absolutely continuous part

$$A_C(u_1, \cdots, u_n) = \int_0^{u_1} \cdots \int_0^{u_n} \frac{\partial^n}{\partial s_1 \cdots \partial s_n} C(s_1, \cdots, s_n) ds_1 \cdots ds_n \tag{8}$$

and a singular part with support on a zero-measure set

$$S_C(u_1, \cdots, u_n) = C(u_1, \cdots, u_n) - A_C(u_1, \cdots, u_n) \tag{9}$$

The absolutely continuous copulas are dense in the set of all copulas.

There are three well-known functions extended from bivariate copulas

$$W^n(\boldsymbol{u}) = \max(u_1 + u_2 + \cdots + u_n - n + 1, 0) \tag{10}$$

$$M^n(\boldsymbol{u}) = \min(u_1, u_2, \cdots, u_n) \tag{11}$$

$$\Pi^n(\boldsymbol{u}) = u_1 u_2 \cdots u_n \tag{12}$$

$W^n$ and $M^n$ are the Fréchet-Hoeffding lower and upper bounds as for any copula $C$,

$$W^n(\boldsymbol{u}) \leq C(\boldsymbol{u}) \leq M^n(\boldsymbol{u}) \tag{13}$$

$W^n$ is not a copula when $n > 2$. $M^n$ is the copula of $X_1, X_2, \dots, X_n$ if and only if each of $X_1, X_2, \dots, X_n$ is almost surely a strictly increasing Borel-measurable function of any of the others, where $\Pi^n$ is a copula of $X_1, X_2, \dots, X_n$ if and only if $X_1, X_2, \dots, X_n$ are independent. $M^n$ is a singular copula while $\Pi^n$ is absolutely continuous with density 1.

We can also define the $k$-dimensional marginal of an $n$-dimensional copula $C$ as

$$C_k(u_1, \cdots, u_k) = C(u_1, \cdots, u_k, 1, \cdots, 1),$$

which is actually a $k$-dimensional copula.

## 4. A new class of multivariate dependence measures

The nonsymmetric bivariate dependence measures defined in Li (2015) are based on first order partial derivative of the bivariate copula function and have the distance-like form

$$\tau(C) = \int_0^1 \int_0^1 \varphi(\partial_1(C(u,v) - \Pi(u,v)))dudv \tag{14}$$

where $\Pi(u,v) = uv$ is the independent copula. Specifically, the two known measures (see Dette, Siburg and Stoimenov, 2010, and Trutschnig, 2011) are

$$\tau_1(C) = 3\int_0^1 \int_0^1 |\partial_1(C(u,v) - \Pi(u,v))|dudv \tag{15}$$

and

$$\tau_2(C) = 6\int_0^1 \int_0^1 (\partial_1(C(u,v) - \Pi(u,v)))^2 dudv \tag{16}$$

For the multivariate case, we will define a similar measure that satisfies the conditions introduced in Sec. 2. For this purpose, we will focus on copulas $C(u_1, \cdots, u_n, v)$ of dimension $n + 1$ and try to measure the dependence of $V$ on $U_1, \cdots, U_n$.

Proposition 4.1. The conditional expectation

$$E(I_{V \le v} | U_1 = u_1, \cdots, U_n = u_n) = \lim_{\Delta u_i \to 0} \frac{E\left(I_{u_1 \le U_1 \le u_1 + \Delta u_1}, \cdots, I_{u_n \le U_n \le u_n + \Delta u_n}, I_{V \le v}\right)}{E\left(I_{u_1 \le U_1 \le u_1 + \Delta u_1}, \cdots, I_{u_n \le U_n \le u_n + \Delta u_n}\right)} \tag{17}$$

is defined almost everywhere (a.e.) and is bounded in the range $[0,1]$.

$$E(I_{V \le v} | U_1 = u_1, \cdots, U_n = u_n) = \frac{\frac{\partial^n}{\partial u_1 \cdots \partial u_n} C(u_1, \cdots, u_n, v)}{\frac{\partial^n}{\partial u_1 \cdots \partial u_n} C(u_1, \cdots, u_n, 1)} \quad \text{a.e.} \tag{18}$$

Proof.

$$E(I_{V\le v}|U_1 = u_1, \cdots, U_n = u_n) = \lim_{\Delta u_i\to 0} \frac{E\left(I_{u_1\le U_1\le u_1+\Delta u_1}, \cdots, I_{u_n\le U_n\le u_n+\Delta u_n}, I_{V\le v}\right)}{E\left(I_{u_1\le U_1\le u_1+\Delta u_1}, \cdots, I_{u_n\le U_n\le u_n+\Delta u_n}\right)}$$

$$= \lim_{\Delta u_i\to 0} \frac{V_C^n(u_1, \cdots, u_n, v; u_1 + \Delta u_1, \cdots, u_n + \Delta u_n, v)}{V_C^n(u_1, \cdots, u_n, 1; u_1 + \Delta u_1, \cdots, u_n + \Delta u_n, 1)}$$

$$= \frac{\frac{\partial^n}{\partial u_1 \cdots \partial u_n} C(u_1, \cdots, u_n, v)}{\frac{\partial^n}{\partial u_1 \cdots \partial u_n} C(u_1, \cdots, u_n, 1)} \qquad \text{a.e.}$$

We have used

$$\lim_{\Delta u_i\to 0} \frac{V_C^n(u_1, \cdots, u_n, v; u_1 + \Delta u_1, \cdots, u_n + \Delta u_n, v)}{\Delta u_1 \cdots \Delta u_n} = \frac{\partial^n}{\partial u_1 \cdots \partial u_n} C(u_1, \cdots, u_n, v) \qquad \text{a.e.}$$

Since

$$V_C^n(u_1, \cdots, u_n, v; u_1 + \Delta u_1, \cdots, u_n + \Delta u_n, v) \le V_C^n(u_1, \cdots, u_n, v; u_1 + \Delta u_1, \cdots, u_n + \Delta u_n, 1)$$

We also have

$$0 \le \frac{\partial^n}{\partial u_1 \cdots \partial u_n} C(u_1, \cdots, u_n, v) \le \frac{\partial^n}{\partial u_1 \cdots \partial u_n} C(u_1, \cdots, u_n, 1) \qquad \text{a.e.}$$

If $\frac{\partial^n}{\partial u_1 \cdots \partial u_n} C(u_1, \cdots, u_n, 1) = 0$, we can always set $\frac{\frac{\partial^n}{\partial u_1 \cdots \partial u_n} C(u_1, \cdots, u_n, v)}{\frac{\partial^n}{\partial u_1 \cdots \partial u_n} C(u_1, \cdots, u_n, 1)} = 0$. Thus,

$$0 \le E(I_{V\le v}|U_1 = u_1, \cdots, U_n = u_n) = \frac{\frac{\partial^n}{\partial u_1 \cdots \partial u_n} C(u_1, \cdots, u_n, v)}{\frac{\partial^n}{\partial u_1 \cdots \partial u_n} C(u_1, \cdots, u_n, 1)} \le 1 \qquad \text{a.e.} \qquad \square$$

Equation (18) was used in Embrechts, Lindskog and McNeil (2003) to generate dependent random variables from a copula. A similar result was proved in Schmitz (2003), Theorem 2.27.

We define the new multivariate dependence measure as follows:

$$\tau(C) = 6 \int_0^1 \cdots \int_0^1 \left(E(I_{V\le v}|U_1 = u_1, \cdots, U_n = u_n) - E(I_{V\le v})\right)^2 dC(u_1, \cdots, u_n, 1) dv$$

$$= 6 \int_0^1 \cdots \int_0^1 \left( \frac{\frac{\partial^n}{\partial u_1 \cdots \partial u_n} C(u_1, \cdots, u_n, v)}{\frac{\partial^n}{\partial u_1 \cdots \partial u_n} C(u_1, \cdots, u_n, 1)} - v \right)^2 \frac{\partial^n}{\partial u_1 \cdots \partial u_n} C(u_1, \cdots, u_n, 1) du_1 \cdots du_n dv \qquad (19)$$

It reduces to the bivariate measure defined in Equation (16) when $n = 1$.

Below we verify the conditions in Sec. 2. For a”), we note that copulas with absolutely continuous density are dense in the set of all copulas, so the measure can be extended to any copulas. Obviously b”) is true as the measure is not symmetric in $V$. For c”), we have

$$\tau(C) = 6\int_0^1 \cdots \int_0^1 \left( \frac{\frac{\partial^n}{\partial u_1 \cdots \partial u_n} C(u_1, \cdots, u_n, v)}{\frac{\partial^n}{\partial u_1 \cdots \partial u_n} C(u_1, \cdots, u_n, 1)} \right)^2 \frac{\partial^n}{\partial u_1 \cdots \partial u_n} C(u_1, \cdots, u_n, 1) du_1 \cdots du_n dv - 2$$

$$\leq 6\int_0^1 \cdots \int_0^1 \left( \frac{\frac{\partial^n}{\partial u_1 \cdots \partial u_n} C(u_1, \cdots, u_n, v)}{\frac{\partial^n}{\partial u_1 \cdots \partial u_n} C(u_1, \cdots, u_n, 1)} \right) \frac{\partial^n}{\partial u_1 \cdots \partial u_n} C(u_1, \cdots, u_n, 1) du_1 \cdots du_n dv - 2$$

$$= 1 \tag{20}$$

So $0 \leq \tau(C) \leq 1$.

If $\tau(C) = 0$, then $\frac{\frac{\partial^n}{\partial u_1 \cdots \partial u_n} C(u_1, \cdots, u_n, v)}{\frac{\partial^n}{\partial u_1 \cdots \partial u_n} C(u_1, \cdots, u_n, 1)} = v$ almost surely. Thus

$$\frac{\partial^n}{\partial u_1 \cdots \partial u_n} C(u_1, \cdots, u_n, v) = \frac{\partial^n}{\partial u_1 \cdots \partial u_n} C(u_1, \cdots, u_n, 1) \cdot v$$

or

$$C(u_1, \cdots, u_n, v) = C(u_1, \cdots, u_n, 1) \cdot v \tag{21}$$

which means that $V$ is independent of $U_1, \cdots, U_n$, even though $U_1, \cdots, U_n$ can still be correlated within themselves, thus d”) holds.

Proposition 4.2. $V$ is almost surely a function of $U_1, \cdots, U_n$ if and only if $\tau(C) = 1$.

Proof. We follow the proof for the bivariate case in Darsow, Nguyen and Olsen (1992), p640. If $\tau(C) = 1$, then, according to Equation (20), we have almost surely

$$E(I_{V \leq v} | U_1 = u_1, \cdots, U_n = u_n) = \frac{\frac{\partial^n}{\partial u_1 \cdots \partial u_n} C(u_1, \cdots, u_n, v)}{\frac{\partial^n}{\partial u_1 \cdots \partial u_n} C(u_1, \cdots, u_n, 1)} \in \{0,1\} \tag{22}$$

Assume the random variables are defined on a probability space $\{\Omega, \Sigma, \mathrm{P}\}$ and measurable function $U: \Omega \rightarrow I^n$ such that $U(\omega) = (u_1, \cdots, u_n)$. For any Borel set $B \in I^n$,

$$\int_{U^{-1}(B)} E(I_{V \leq v} | U_1, \cdots, U_n) I_{V \leq v} \, dP$$

$$= \int_{U^{-1}(B)} E[E(I_{V \leq v} | U_1, \cdots, U_n) I_{V \leq v} | U_1, \cdots, U_n] \, dP$$

$$= \int_{U^{-1}(B)} E(I_{V \leq v} | U_1, \cdots, U_n)^2 \, dP$$

$$= \int_{U^{-1}(B)} E(I_{V\le v}|U_1,\cdots,U_n)\, dP$$

$$= \int_{U^{-1}(B)} I_{V\le v}\, dP = \int_{U^{-1}(B)} {I_{V\le v}}^2\, dP$$

It follows

$$0 = \int_{U^{-1}(B)} (I_{V\le v} - E(I_{V\le v}|U_1,\cdots,U_n))^2\, dP$$

which implies $I_{V\le v} = E(I_{V\le v}|U_1,\cdots,U_n)$ almost surely for all $v$. Thus $V$ is measurable with respect to the $\sigma$ algebra generated by $U_1,\cdots,U_n$ or $V$ is almost surely a function of $U_1,\cdots,U_n$.

Conversely, if $V$ is a function of $U_1,\cdots,U_n$, then the $\sigma$ algebra generated by $V$ is contained in the $\sigma$ algebra generated by $U_1,\cdots,U_n$. So that $I_{V\le v} = E(I_{V\le v}|U_1,\cdots,U_n)$ almost surely, which implies $\tau(C) = 1$. □

The proof of condition f”) will be discussed in the next section.

## 5. Generalized $*$ product and DPI condition

The $*$ product for bivariate copulas $A$ and $B$ (Darsow, Nguyen and Olsen, 1992) is useful in describing Markov process through copula and is defined as follows,

$$(A*B)(u,v) = \int_0^1 \partial_2 A(u,t)\cdot \partial_1 B(t,v)dt \tag{23}$$

It has been used in a previous paper (Li, 2015) to prove condition f’). We would like to extend it to higher dimensions and use it to prove condition f”) for the multivariate case.

Let $A \in \mathbb{C}^{2n}$, $B \in \mathbb{C}^{n+1}$ and $D \in \mathbb{C}^n$ such that

$$A(1,\cdots,1,s_1,\cdots,s_n) = B(s_1,\cdots,s_n,1) = D(s_1,\cdots,s_n). \tag{24}$$

We define

$$(A*B)(u_1,\cdots,u_n,v) = \int_0^1 \cdots \int_0^1 \frac{\frac{\partial^n}{\partial s_1\cdots\partial s_n} A(u_1,\cdots,u_n,s_1,\cdots,s_n)}{\frac{\partial^n}{\partial s_1\cdots\partial s_n} A(1,\cdots,1,s_1,\cdots,s_n)}\cdot \frac{\frac{\partial^n}{\partial s_1\cdots\partial s_n} B(s_1,\cdots,s_n,v)}{\frac{\partial^n}{\partial s_1\cdots\partial s_n} B(s_1,\cdots,s_n,1)}\cdot \frac{\partial^n}{\partial s_1\cdots\partial s_n} D(s_1,\cdots,s_n) ds_1\cdots ds_n \tag{25}$$

The first two terms in the integrand are bounded in [0,1] using the same argument in Proposition 4.1. It is easy to verify that $(A*B)(u_1,\cdots,u_n,v)$ satisfies the first two conditions of the copula definition. To verify the third condition, we have

$$V_{A*B}(\boldsymbol{u_1}, v_1; \boldsymbol{u_2}, v_2) = \int_0^1 \cdots \int_0^1 \frac{\frac{\partial^n}{\partial s_1 \cdots \partial s_n} V_A^n(\boldsymbol{u_1}, \boldsymbol{s}; \boldsymbol{u_2}, \boldsymbol{s})}{\frac{\partial^n}{\partial s_1 \cdots \partial s_n} A(1, \cdots, 1, s_1, \cdots, s_n)} \cdot \frac{\frac{\partial^n}{\partial s_1 \cdots \partial s_n} V_B^1(\boldsymbol{s}, v_1; \boldsymbol{s}, v_2)}{\frac{\partial^n}{\partial s_1 \cdots \partial s_n} B(s_1, \cdots, s_n, 1)} \cdot \frac{\partial^n}{\partial s_1 \cdots \partial s_n} D(s_1, \cdots, s_n) ds_1 \cdots ds_n \tag{26}$$

where $\boldsymbol{s} = (s_1, \cdots, s_n)$, volume $V_{A*B}$ is defined on an (n+1)-box $[\boldsymbol{u_1}, v_1; \boldsymbol{u_2}, v_2]$, volume $V_A^n(\boldsymbol{u_1}, \boldsymbol{s}; \boldsymbol{u_2}, \boldsymbol{s})$ is defined on an n-box $[\boldsymbol{u_1}, \boldsymbol{s}; \boldsymbol{u_2}, \boldsymbol{s}]$ in 2n space, and volume $V_B^1(\boldsymbol{s}, v_1; \boldsymbol{s}, v_2)$ is defined on a 1-box $[\boldsymbol{s}, v_1; \boldsymbol{s}, v_2]$ in n+1 space. As each term in the integration is non-negative, $V_{A*B}(\boldsymbol{u_1}, v_1; \boldsymbol{u_2}, v_2)$ is also non-negative. Thus $A * B$ is indeed a copula.

Note that $(A * B)(u_1, \cdots, u_n, 1) = A(u_1, \cdots, u_n, 1, \cdots, 1)$, which is just the copula of $U_1, \cdots, U_n$. Besides, if $B(s_1, \cdots, s_n, v) = B(s_1, \cdots, s_n, 1) \cdot v$ or $V$ is independent of $S_1, \cdots, S_n$, then $(A * B)(u_1, \cdots, u_n, v) = A(u_1, \cdots, u_n, 1, \cdots, 1) \cdot v$, which means $V$ is also independent of $U_1, \cdots, U_n$. If $A(u_1, \cdots, u_n, s_1, \cdots, s_n) = A(u_1, \cdots, u_n, 1, \cdots, 1) \cdot A(1, \cdots, 1,, s_1, \cdots, s_n)$, then we also have $(A * B)(u_1, \cdots, u_n, v) = A(u_1, \cdots, u_n, 1, \cdots, 1) \cdot v$ or $V$ is again independent of $U_1, \cdots, U_n$.

PROPOSITION 5.1 If random variables $U_1, \cdots, U_n$ and $V$ are conditionally independent given $S_1, \cdots, S_n$, then

$$C(u_1, \cdots, u_n, v) = A(u_1, \cdots, u_n, s_1, \cdots, s_n) * B(s_1, \cdots, s_n, v) \tag{27}$$

Proof. The conditional independence means

$$\begin{aligned} &E\left(I_{U_1 \le u_1} \cdots I_{U_n \le u_n} I_{V \le v} | S_1 = s_1, \cdots, S_n = s_n\right) \\ &= E\left(I_{U_1 \le u_1} \cdots I_{U_n \le u_n} | S_1 = s_1, \cdots, S_n = s_n\right) \cdot E(I_{V \le v} | S_1 = s_1, \cdots, S_n = s_n) \end{aligned} \tag{28}$$

Note that

$$E\left(I_{U_1 \le u_1} \cdots I_{U_n \le u_n} | S_1 = s_1, \cdots, S_n = s_n\right) = \frac{\frac{\partial^n}{\partial s_1 \cdots \partial s_n} A(u_1, \cdots, u_n, s_1, \cdots, s_n)}{\frac{\partial^n}{\partial s_1 \cdots \partial s_n} A(1, \cdots, 1, s_1, \cdots, s_n)} \quad \text{a.e.} \tag{29}$$

and

$$E(I_{V \le v} | S_1 = s_1, \cdots, S_n = s_n) = \frac{\frac{\partial^n}{\partial s_1 \cdots \partial s_n} B(s_1, \cdots, s_n, v)}{\frac{\partial^n}{\partial s_1 \cdots \partial s_n} B(s_1, \cdots, s_n, 1)} \quad \text{a.e.} \tag{30}$$

Integrating both sides of Equation (28) on $S_1, \cdots, S_n$ leads to Equation (27). □

This is similar to the bivariate case in Darsow, Nguyen and Olsen (1992), p610. We can say that $(U_1, \cdots, U_n)$, $(S_1, \cdots, S_n)$, $V$ form a Markov chain. $V$ is less dependent on $U_1, \cdots, U_n$ than on $S_1, \cdots, S_n$ as the dependence of $V$ on $U_1, \cdots, U_n$ is through $S_1, \cdots, S_n$. This should be reflected in the dependence measure.

A general multivariate dependence measure can be defined as

$$\tau(C) = \int_0^1 \cdots \int_0^1 \varphi\left(\frac{\frac{\partial^n}{\partial u_1 \cdots \partial u_n} C(u_1, \cdots, u_n, v)}{\frac{\partial^n}{\partial u_1 \cdots \partial u_n} C(u_1, \cdots, u_n, 1)} - v\right) \frac{\partial^n}{\partial u_1 \cdots \partial u_n} C(u_1, \cdots, u_n, 1) du_1 \cdots du_n dv \tag{31}$$

where $\varphi$ is a convex function and is not explicitly dependent on $u_1, \cdots, u_n$.

PROPOSITION 5.2. If $(U_1, \cdots, U_n)$, $(S_1, \cdots, S_n)$, $V$ form a Markov chain, then

$$\tau\big(C(u_1, \cdots, u_n, v)\big) \leq \tau\big(B(s_1, \cdots, s_n, v)\big) \tag{32}$$

PROOF: It suffices to consider that the copula $A(u_1, \cdots, u_n, s_1, \cdots, s_n)$ is absolutely continuous as absolutely continuous copulas are dense in the set of all copulas. Using Jensen's inequality,

$$\begin{aligned}
\tau\big(C(u_1, \cdots, u_n, v)\big) &= \tau\big(A(u_1, \cdots, u_n, s_1, \cdots, s_n) * B(s_1, \cdots, s_n, v)\big) \\
&= \int_0^1 \cdots \int_0^1 \varphi\left(\frac{\frac{\partial^n}{\partial u_1 \cdots \partial u_n}\big(A(u_1, \cdots, u_n, s_1, \cdots, s_n) * B(s_1, \cdots, s_n, v)\big)}{\frac{\partial^n}{\partial u_1 \cdots \partial u_n} C(u_1, \cdots, u_n, 1)} - v\right) \cdot \\
&\qquad \frac{\partial^n}{\partial u_1 \cdots \partial u_n} C(u_1, \cdots, u_n, 1) du_1 \cdots du_n dv
\end{aligned}$$

$$= \int_0^1 \cdots \int_0^1 \varphi\left(\frac{\frac{\partial^n}{\partial u_1 \cdots \partial u_n}}{\frac{\partial^n}{\partial u_1 \cdots \partial u_n} C(u_1, \cdots, u_n, 1)} \int_0^1 \cdots \int_0^1 \frac{\frac{\partial^n}{\partial s_1 \cdots \partial s_n} A(u_1, \cdots, u_n, s_1, \cdots, s_n)}{\frac{\partial^n}{\partial s_1 \cdots \partial s_n} A(1, \cdots, 1, s_1, \cdots, s_n)} \cdot \frac{\frac{\partial^n}{\partial s_1 \cdots \partial s_n} B(s_1, \cdots, s_n, v)}{\frac{\partial^n}{\partial s_1 \cdots \partial s_n} B(s_1, \cdots, s_n, 1)} \cdot \frac{\partial^n}{\partial s_1 \cdots \partial s_n} D(s_1, \cdots, s_n) ds_1 \cdots ds_n - v\right) \cdot \frac{\partial^n}{\partial u_1 \cdots \partial u_n} C(u_1, \cdots, u_n, 1) du_1 \cdots du_n dv$$

$$= \int_0^1 \cdots \int_0^1 \varphi\left(\int_0^1 \cdots \int_0^1 \frac{\frac{\partial^{2n}}{\partial u_1 \cdots \partial u_n \partial s_1 \cdots \partial s_n} A(u_1, \cdots, u_n, s_1, \cdots, s_n)}{\frac{\partial^n}{\partial u_1 \cdots \partial u_n} C(u_1, \cdots, u_n, 1) \cdot \frac{\partial^n}{\partial s_1 \cdots \partial s_n} A(1, \cdots, 1, s_1, \cdots, s_n)} \cdot \left(\frac{\frac{\partial^n}{\partial s_1 \cdots \partial s_n} B(s_1, \cdots, s_n, v)}{\frac{\partial^n}{\partial s_1 \cdots \partial s_n} B(s_1, \cdots, s_n, 1)} - v\right) \cdot \frac{\partial^n}{\partial s_1 \cdots \partial s_n} D(s_1, \cdots, s_n) ds_1 \cdots ds_n\right) \frac{\partial^n}{\partial u_1 \cdots \partial u_n} C(u_1, \cdots, u_n, 1) du_1 \cdots du_n dv$$

$$
\begin{aligned}
\le \quad & \int_0^1 \cdots \int_0^1 \left( \int_0^1 \cdots \int_0^1 \frac{\frac{\partial^{2n}}{\partial u_1 \cdots \partial u_n \partial s_1 \cdots \partial s_n} A(u_1, \cdots, u_n, s_1, \cdots, s_n)}{\frac{\partial^n}{\partial u_1 \cdots \partial u_n} C(u_1, \cdots, u_n, 1) \cdot \frac{\partial^n}{\partial s_1 \cdots \partial s_n} A(1, \cdots, 1, s_1, \cdots, s_n)} \cdot \right. \\
& \left. \varphi\left( \frac{\frac{\partial^n}{\partial s_1 \cdots \partial s_n} B(s_1, \cdots, s_n, v)}{\frac{\partial^n}{\partial s_1 \cdots \partial s_n} B(s_1, \cdots, s_n, 1)} - v \right) \cdot \frac{\partial^n}{\partial s_1 \cdots \partial s_n} D(s_1, \cdots, s_n) ds_1 \cdots ds_n \right) \cdot \\
& \frac{\partial^n}{\partial u_1 \cdots \partial u_n} C(u_1, \cdots, u_n, 1) du_1 \cdots du_n dv \\
= \quad & \int_0^1 \cdots \int_0^1 \left( \int_0^1 \cdots \int_0^1 \frac{\frac{\partial^{2n}}{\partial u_1 \cdots \partial u_n \partial s_1 \cdots \partial s_n} A(u_1, \cdots, u_n, s_1, \cdots, s_n)}{\frac{\partial^n}{\partial u_1 \cdots \partial u_n} C(u_1, \cdots, u_n, 1) \cdot \frac{\partial^n}{\partial s_1 \cdots \partial s_n} A(1, \cdots, 1, s_1, \cdots, s_n)} \cdot \right. \\
& \left. \frac{\partial^n}{\partial u_1 \cdots \partial u_n} C(u_1, \cdots, u_n, 1) du_1 \cdots du_n \right) \cdot \\
& \varphi\left( \frac{\frac{\partial^n}{\partial s_1 \cdots \partial s_n} B(s_1, \cdots, s_n, v)}{\frac{\partial^n}{\partial s_1 \cdots \partial s_n} B(s_1, \cdots, s_n, 1)} - v \right) \frac{\partial^n}{\partial s_1 \cdots \partial s_n} D(s_1, \cdots, s_n) ds_1 \cdots ds_n dv \\
= \quad & \int_0^1 \cdots \int_0^1 \varphi\left( \frac{\frac{\partial^n}{\partial s_1 \cdots \partial s_n} B(s_1, \cdots, s_n, v)}{\frac{\partial^n}{\partial s_1 \cdots \partial s_n} B(s_1, \cdots, s_n, 1)} - v \right) \cdot \\
& \frac{\partial^n}{\partial s_1 \cdots \partial s_n} B(s_1, \cdots, s_n, 1) ds_1 \cdots ds_n dv \\
= \quad & \tau\big(B(s_1, \cdots, s_n, v)\big) \qquad (33)
\end{aligned}
$$

where we have used the following

$$
\int_0^1 \cdots \int_0^1 \frac{\frac{\partial^{2n}}{\partial u_1 \cdots \partial u_n \partial s_1 \cdots \partial s_n} A(u_1, \cdots, u_n, s_1, \cdots, s_n)}{\frac{\partial^n}{\partial u_1 \cdots \partial u_n} C(u_1, \cdots, u_n, 1) \cdot \frac{\partial^n}{\partial s_1 \cdots \partial s_n} A(1, \cdots, 1, s_1, \cdots, s_n)} \frac{\partial^n}{\partial u_1 \cdots \partial u_n} C(u_1, \cdots, u_n, 1) du_1 \cdots du_n = 1 \qquad (34)
$$

and

$$
\int_0^1 \cdots \int_0^1 \frac{\frac{\partial^{2n}}{\partial u_1 \cdots \partial u_n \partial s_1 \cdots \partial s_n} A(u_1, \cdots, u_n, s_1, \cdots, s_n)}{\frac{\partial^n}{\partial u_1 \cdots \partial u_n} C(u_1, \cdots, u_n, 1) \cdot \frac{\partial^n}{\partial s_1 \cdots \partial s_n} A(1, \cdots, 1, s_1, \cdots, s_n)} \frac{\partial^n}{\partial s_1 \cdots \partial s_n} D(s_1, \cdots, s_n) ds_1 \cdots ds_n = 1 \qquad (35)
$$

In Equation (33), the first line uses Proposition 5.1, the second line uses Equation (31), the third line plugs in Equation (25), the fourth line uses Equation (35) to move $v$ inside, the fifth line uses Jensen's inequality together with Equation (35), the sixth line uses Fubini's Theorem to integrate out $u_i$'s first, the seventh line is the result of using Equation (34) and (24), which leads to the final result according to Equation (31).

If $\varphi$ is strictly convex, then the equal sign holds if $\frac{\frac{\partial^n}{\partial s_1 \cdots \partial s_n} B(s_1, \cdots, s_n, v)}{\frac{\partial^n}{\partial s_1 \cdots \partial s_n} B(s_1, \cdots, s_n, 1)} - v$ is almost constant in $s_1, \cdots, s_n$ with respect to the measure defined by the following density

$$f(s_1,\cdots,s_n;u_1,\cdots,u_n,v) = \frac{\frac{\partial^{2n}}{\partial u_1\cdots\partial u_n\partial s_1\cdots\partial s_n}A(u_1,\cdots,u_n,s_1,\cdots,s_n)}{\frac{\partial^n}{\partial u_1\cdots\partial u_n}C(u_1,\cdots,u_n,1)\cdot\frac{\partial^n}{\partial s_1\cdots\partial s_n}A(1,\cdots,1,s_1,\cdots,s_n)}\frac{\partial^n}{\partial s_1\cdots\partial s_n}D(s_1,\cdots,s_n) \tag{36}$$

on $s_1,\cdots,s_n \in I^n$ for almost all $(u_1,\cdots,u_n) \in I^n, v \in I$. □

This is a generalized form of the Data Processing Inequality (DPI) in information theory, see, e. g., Chapter 2 of Cover and Thomas (1991). DPI says that if $X$, $Y$, $Z$ form a Markov chain, then $MI(X,Y) \geq MI(X,Z)$ for Shannon's mutual information. It implies that no processing of $Y$ can increase the information that $Y$ contains about $X$. A more general form of DPI for symmetric bivariate dependence measures was discussed in Kinney and Atwal (2014). Condition f") is a nonsymmetric multivariate extension to the self-equitable property defined in their paper, where the nonsymmetric bivariate case has already been discussed in Li (2015). The concept of equitability was initially introduced by Reshef, et al (2011), which was used to characterize a dependence measure that roughly equals the coefficient of determination ($R^2$) relative to various (nonlinear) regression functions. Self-equitability is a related but more general concept. In Kinney and Atwal (2014), self-equitability was defined for symmetric measures. It was pointed out by Murrell, Murrell and Murrell (2014) that regression-based equitability should be defined for nonsymmetric measures. Next we define self-equitability for our nonsymmetric multivariate dependence measures.

Definition 5.1. A dependence measure $R(X_1,X_2,\dots,X_n,Y)$ is self-equitable if and only if it satisfies $R(g(X_1,X_2,\dots,X_n),Y) = R(X_1,X_2,\dots,X_n,Y)$ whenever $g: \mathbb{R}^n \to \mathbb{R}^n$ is a deterministic function and $(X_1,X_2,\dots,X_n)$, $g(X_1,X_2,\dots,X_n)$ and $Y$ form a Markov chain.

PROPOSITION 5.3 If $f$ is a continuous bijection on $\mathbb{R}^n$, then $\tau\big(C_{f(X_1,X_2,\dots,X_n)Y}\big) = \tau\big(C_{X_1,X_2,\dots,X_n,Y}\big)$.

PROOF: As $f$ is a bijective mapping, $(X_1,X_2,\dots,X_n)$, $f(X_1,X_2,\dots,X_n)$, $Y$ form a Markov chain. Thus $\tau\big(C_{X_1,X_2,\dots,X_n,Y}\big) \leq \tau\big(C_{f(X_1,X_2,\dots,X_n)Y}\big)$. On the other hand, for any continuous mapping $f: \mathbb{R}^n \to \mathbb{R}^n$ , $f(X_1,X_2,\dots,X_n)$, $(X_1,X_2,\dots,X_n)$, $Y$ also form a Markov chain, which implies $\tau\big(C_{f(X_1,X_2,\dots,X_n)Y}\big) \leq \tau\big(C_{X_1,X_2,\dots,X_n,Y}\big)$. Therefore $\tau\big(C_{f(X_1,X_2,\dots,X_n)Y}\big) = \tau\big(C_{X_1,X_2,\dots,X_n,Y}\big)$. □

From the proof of proposition 5.3, it is obvious that the measure is also self-equitable.

PROPOSITION 5.4 If $g$ is a continuous strictly monotonic transformation on $\mathbb{R}$, then $\tau\big(C_{X_1,X_2,\dots,X_n,g(Y)}\big) = \tau\big(C_{X_1,X_2,\dots,X_n,Y}\big)$.

PROOF: If $g$ is strictly increasing, then $C_{X_1,X_2,\dots,X_n,g(Y)} = C_{X_1,X_2,\dots,X_n,Y}$. If $g$ is strictly decreasing, $C_{X_1,X_2,\dots,X_n,g(Y)}(u_1,\cdots,u_n,v) = C_{X_1,X_2,\dots,X_n}(u_1,\cdots,u_n) - C_{X_1,X_2,\dots,X_n,Y}(u_1,\cdots,u_n,1-v)$, see e.g. Embrechts, Lindskog and McNeil (2003). Therefore, $C_{X_1,X_2,\dots,X_n,g(Y)}(u_1,\cdots,u_n,v) -$ $C_{X_1,X_2,\dots,X_n}(u_1,\cdots,u_n)\cdot v = C_{X_1,X_2,\dots,X_n}(u_1,\cdots,u_n)\cdot(1-v) - C_{X_1,X_2,\dots,X_n,Y}(u_1,\cdots,u_n,1-v)$, which, by change of variable, leads to $\tau\left(C_{X_1,X_2,\dots,X_n,g(Y)}\right) = \tau\left(C_{X_1,X_2,\dots,X_n,Y}\right)$ as long as the measure is a symmetric function of $C_{X_1,X_2,\dots,X_n,Y} - C_{X_1,X_2,\dots,X_n}(u_1,\cdots,u_n)\cdot v$, which is generally true for distance-like measures. □

Therefore, for certain general form of dependence measure in Equation (31) with convex function $\varphi$, we have proven a stronger condition than f").

> f"') If $f$ is a continuous bijection on $\mathbb{R}^n$ and $g$ is a strictly monotonic transformation on $\mathbb{R}$, then $R(f(X_1,X_2,\dots,X_n),g(Y)) = R(X_1,X_2,\dots,X_n,Y)$.

The general form of multivariate dependence measures in distance form would be like the following

$$\tau_\alpha(C) = \int_0^1\cdots\int_0^1 \left|\frac{\frac{\partial^n}{\partial u_1\cdots\partial u_n}C(u_1,\cdots,u_n,v)}{\frac{\partial^n}{\partial u_1\cdots\partial u_n}C(u_1,\cdots,u_n,1)} - v\right|^\alpha \frac{\partial^n}{\partial u_1\cdots\partial u_n}C(u_1,\cdots,u_n,1)du_1\cdots du_n dv \tag{37}$$

for $\alpha \geq 1$.

The dependence measure may also be written in entropy form similar to Rényi's mutual information (Rényi, 1961 and Li, 2015),

$$R_\alpha(C) = \frac{1}{\alpha-1}log\left(\int_0^1\cdots\int_0^1\left(\frac{\frac{\partial^n}{\partial u_1\cdots\partial u_n}C(u_1,\cdots,u_n,v)}{v\cdot\frac{\partial^n}{\partial u_1\cdots\partial u_n}C(u_1,\cdots,u_n,1)}\right)^\alpha \cdot \frac{\partial^n}{\partial u_1\cdots\partial u_n}C(u_1,\cdots,u_n,1)du_1\cdots du_n dv\right) \tag{38}$$

where $0 < \alpha < 2$. In the limit $\alpha \to 1$, it reduces to

$$R(C) = \int_0^1\cdots\int_0^1 \frac{\frac{\partial^n}{\partial u_1\cdots\partial u_n}C(u_1,\cdots,u_n,v)}{v\cdot\frac{\partial^n}{\partial u_1\cdots\partial u_n}C(u_1,\cdots,u_n,1)}\cdot \log\left(\frac{\frac{\partial^n}{\partial u_1\cdots\partial u_n}C(u_1,\cdots,u_n,v)}{v\cdot\frac{\partial^n}{\partial u_1\cdots\partial u_n}C(u_1,\cdots,u_n,1)}\right)\frac{\partial^n}{\partial u_1\cdots\partial u_n}C(u_1,\cdots,u_n,1)du_1\cdots du_n dv \tag{39}$$

In the 3-dimensional case of $R(X_1,X_2,Y)$, the dependence measure of Equation (19) takes the form

$$\tau(C) = 6\int_0^1\int_0^1\int_0^1\left(\frac{\frac{\partial^2}{\partial u_1\partial u_2}C(u_1,u_2,v)}{\frac{\partial^2}{\partial u_1\partial u_2}C(u_1,u_2,1)} - v\right)^2 \frac{\partial^2}{\partial u_1\partial u_2}C(u_1,u_2,1)du_1du_2dv \tag{40}$$

Equation (39) for $n = 2$ becomes

$$R(C)=\int_0^1\int_0^1\int_0^1 \frac{\frac{\partial^2}{\partial u_1\partial u_2}C(u_1,u_2,v)}{v\cdot\frac{\partial^2}{\partial u_1\partial u_2}C(u_1,u_2,1)}\log\left(\frac{\frac{\partial^2}{\partial u_1\partial u_2}C(u_1,u_2,v)}{v\cdot\frac{\partial^2}{\partial u_1\partial u_2}C(u_1,u_2,1)}\right)\frac{\partial^2}{\partial u_1\partial u_2}C(u_1,u_2,1)du_1du_2dv \tag{40}$$

This can be compared with Shannon's mutual information which is symmetric:

$$MI(C)=\int_0^1\int_0^1\int_0^1 c(u_1,u_2,v)log\big(c(u_1,u_2,v)\big)du_1du_2dv \tag{41}$$

where $c(u_1,u_2,v)=\frac{\partial^3}{\partial u_1\partial u_2\partial v}C(u_1,u_2,v)$ is the copula density. The mutual information would be infinity if there is any singularity in $c(u_1,u_2,v)$, which may not correspond to functional relationship between any variables. This kind of issues for symmetric measures has been discussed in Li (2015).

As an example, let $C(u_1,u_2,v)=\min(u_1,u_2,v)$, which is the Fréchet-Hoeffding upper bound for 3-dimensional copula and is a singular copula with $X_1,X_2,Y$ monotonically increasing function of each other. It is easy to verify that $\tau(C)=R(C)=1$ and $MI(C)=\infty$.

## 6. Further extension

A natural extension of the dependence measure discussed above would be to consider how a group of random variables depends on another group of random variables with the two extremes of independence between the two groups and complete dependence of the first group on the second group. For this purpose, we extend the conditional expectation in Proposition 4.1 to the multivariate case

$$E(I_{V_1\le v_1}\cdots I_{V_m\le v_m}|U_1=u_1,\cdots,U_n=u_n)=\frac{\frac{\partial^n}{\partial u_1\cdots\partial u_n}C(u_1,\cdots,u_n,v_1,\cdots,v_m)}{\frac{\partial^n}{\partial u_1\cdots\partial u_n}C(u_1,\cdots,u_n,1,\cdots,1)} \tag{42}$$

which is again defined almost everywhere and is bounded in the range [0,1]. This kind of extension was also discussed in Schmitz (2003), Corollary 2.28. Similar to Equation (19), we define the quadratic dependence measure as

$$\tau(C)=6\int_0^1\cdots\int_0^1\Big(E\big(I_{V_1\le v_1}\cdots I_{V_m\le v_m}|U_1=u_1,\cdots,U_n=u_n\big) - E\big(I_{V_1\le v_1}\cdots I_{V_m\le v_m}\big)\Big)^2 dC(u_1,\cdots,u_n)\cdot dC(v_1,\cdots,v_m)$$

$$= 6 \int_0^1 \cdots \int_0^1 \left( \frac{\frac{\partial^n}{\partial u_1 \cdots \partial u_n} C(u_1, \cdots, u_n, v_1, \cdots, v_m)}{\frac{\partial^n}{\partial u_1 \cdots \partial u_n} C(u_1, \cdots, u_n)} - C(v_1, \cdots, v_m) \right)^2$$

$$\cdot \frac{\partial^n}{\partial u_1 \cdots \partial u_n} C(u_1, \cdots, u_n) du_1 \cdots du_n \cdot \frac{\partial^n}{\partial v_1 \cdots \partial v_m} C(v_1, \cdots, v_m) dv_1 \cdots dv_m \tag{43}$$

We have simplified the notation of the marginal copulas in Equation (43) as

$$C(u_1, \cdots, u_n, 1, \cdots, 1) = C(u_1, \cdots, u_n)$$

and

$$C(1, \cdots, 1, v_1, \cdots, v_m) = C(v_1, \cdots, v_m).$$

Again, $\tau(C) = 0$ if and only if

$$C(u_1, \cdots, u_n, v_1, \cdots, v_m) = C(u_1, \cdots, u_n) \cdot C(v_1, \cdots, v_m) \tag{44}$$

such that the two groups $(U_1, \cdots, U_n)$ and $(V_1, \cdots, V_m)$ are independent of each other.

However, unlike the previous case, the maximum value of $\tau(C)$ depends on the marginal copula $C(v_1, \cdots, v_m)$ for $m > 1$.

$$\tau(C) = 6 \int_0^1 \cdots \int_0^1 \left( \frac{\frac{\partial^n}{\partial u_1 \cdots \partial u_n} C(u_1, \cdots, u_n, v_1, \cdots, v_m)}{\frac{\partial^n}{\partial u_1 \cdots \partial u_n} C(u_1, \cdots, u_n)} \right)^2$$

$$\cdot \frac{\partial^n}{\partial u_1 \cdots \partial u_n} C(u_1, \cdots, u_n) du_1 \cdots du_n \cdot \frac{\partial^n}{\partial v_1 \cdots \partial v_m} C(v_1, \cdots, v_m) dv_1 \cdots dv_m$$

$$- 6 \int_0^1 \cdots \int_0^1 C(v_1, \cdots, v_m)^2 \frac{\partial^n}{\partial v_1 \cdots \partial v_m} C(v_1, \cdots, v_m) dv_1 \cdots dv_m$$

$$\leq 6 \int_0^1 \cdots \int_0^1 \left( \frac{\frac{\partial^n}{\partial u_1 \cdots \partial u_n} C(u_1, \cdots, u_n, v_1, \cdots, v_m)}{\frac{\partial^n}{\partial u_1 \cdots \partial u_n} C(u_1, \cdots, u_n)} \right)$$

$$\cdot \frac{\partial^n}{\partial u_1 \cdots \partial u_n} C(u_1, \cdots, u_n) du_1 \cdots du_n \cdot \frac{\partial^n}{\partial v_1 \cdots \partial v_m} C(v_1, \cdots, v_m) dv_1 \cdots dv_m$$

$$- 6 \int_0^1 \cdots \int_0^1 C(v_1, \cdots, v_m)^2 \frac{\partial^n}{\partial v_1 \cdots \partial v_m} C(v_1, \cdots, v_m) dv_1 \cdots dv_m$$

$$= 6 \int_0^1 \cdots \int_0^1 (C(v_1, \cdots, v_m) - C(v_1, \cdots, v_m)^2) dC(v_1, \cdots, v_m) \quad (45)$$

The equal sign holds if and only if, almost surely,

$$E(I_{V_1 \le v_1} \cdots I_{V_m \le v_m} | U_1 = u_1, \cdots, U_n = u_n) = \frac{\frac{\partial^n}{\partial u_1 \cdots \partial u_n} C(u_1, \cdots, u_n, v_1, \cdots, v_m)}{\frac{\partial^n}{\partial u_1 \cdots \partial u_n} C(u_1, \cdots, u_n)} \in \{0,1\} \quad (46)$$

This leads to the conclusion that $V_1, \cdots, V_m$ are functions of $U_1, \cdots, U_n$, following the same kind of arguments as in the proof of proposition 4.2.

The maximum value of $\tau(C)$ in Equation (45) can be re-written in terms of the Kendall distribution function as defined in Genest and Rivest (1993, 2001) and Nelson, Quesada Molina, Rodríguez Lallena and Úbeda Flores (2003),

$$K_C(t) = P(C(V_1, \cdots, V_m) \le t), \quad t \in [0,1] \quad (47)$$

where $C(V_1, \cdots, V_m)$ as a function of $V_1, \cdots, V_m$ is a random variable and $K_C(t)$ is its cumulative distribution function. Then the maximum value of $\tau(C)$ will be

$$M\big(C(V_1, \cdots, V_m)\big) = 6 \int_0^1 \cdots \int_0^1 (C(v_1, \cdots, v_m) - C(v_1, \cdots, v_m)^2) dC(v_1, \cdots, v_m)$$

$$= 6 \int_0^1 (t - t^2) dK_C(t) \quad (48)$$

In the special case of $m = 1$ where the copula degenerates for one variable and $K_C(t) = t$, it has constant value of 1 as shown in Equation (20). The Kendall distribution function has already been used for the calculation of Kendall's tau. Now it can be used to calculate bound on the new nonsymmetric dependence measures.

If $(U_1, \cdots, U_n)$, $(S_1, \cdots, S_n)$, $(V_1, \cdots, V_m)$ form a Markov chain or the two groups $(U_1, \cdots, U_n)$, $(V_1, \cdots, V_m)$ are independent conditional on $(S_1, \cdots, S_n)$, then, extending Equations (25) and (27),

$$C(u_1, \cdots, u_n, v_1, \cdots, v_m) = A(u_1, \cdots, u_n, s_1, \cdots, s_n) * B(s_1, \cdots, s_n, v_1, \cdots, v_m)$$

$$= \int_0^1 \cdots \int_0^1 \frac{\frac{\partial^n}{\partial s_1 \cdots \partial s_n} A(u_1, \cdots, u_n, s_1, \cdots, s_n)}{\frac{\partial^n}{\partial s_1 \cdots \partial s_n} A(1, \cdots, 1, s_1, \cdots, s_n)} \cdot \frac{\frac{\partial^n}{\partial s_1 \cdots \partial s_n} B(s_1, \cdots, s_n, v_1, \cdots, v_m)}{\frac{\partial^n}{\partial s_1 \cdots \partial s_n} B(s_1, \cdots, s_n, 1, \cdots, 1)} \cdot \frac{\partial^n}{\partial s_1 \cdots \partial s_n} D(s_1, \cdots, s_n) ds_1 \cdots ds_n$$

(49)

Again, a general multivariate dependence measure can be defined as

$$\tau(C) = \int_0^1 \cdots \int_0^1 \varphi\left( \frac{\frac{\partial^n}{\partial u_1 \cdots \partial u_n} C(u_1, \cdots, u_n, v_1, \cdots, v_m)}{\frac{\partial^n}{\partial u_1 \cdots \partial u_n} C(u_1, \cdots, u_n)} - C(v_1, \cdots, v_m) \right)$$

$$\cdot \frac{\partial^n}{\partial u_1 \cdots \partial u_n} C(u_1, \cdots, u_n) du_1 \cdots du_n \cdot \frac{\partial^n}{\partial v_1 \cdots \partial v_m} C(v_1, \cdots, v_m) dv_1 \cdots dv_m \qquad (50)$$

where $\varphi$ is a convex function and is not explicitly dependent on $u_1, \cdots, u_n$. Proposition 5.2 can be extended with similar arguments,

$$\tau\big(C(u_1, \cdots, u_n, v_1, \cdots, v_m)\big) \le \tau\big(B(s_1, \cdots, s_n, v_1, \cdots, v_m)\big) \qquad (51)$$

So the DPI condition still holds. Thus the dependence measure $R(X_1, \ldots, X_n, Y_1, \ldots, Y_m)$ is invariant under bijective transformations on $(X_1, \ldots, X_n)$ and monotonically increasing transformations on each of $Y_1, \ldots, Y_m$. The dependence measure may not be invariant under strictly decreasing transformations on any of $Y_1, \ldots, Y_m$ if $m > 1$, which is related to the fact that Equation (48) may not be invariant under strictly decreasing transformations. The entropy form similar to Equation (38) can also be readily written down by substituting $v$ with $C(v_1, \cdots, v_m)$.

We remark that, in Sanchez and Trutschnig (2014), for the dependence of $(V_1, \cdots, V_m)$ on $U$, the dependence measure was initially defined as distance from the independent copula $u \cdot v_1 \cdot \cdots \cdot v_m$ instead of $u \cdot C(v_1, \cdots, v_m)$, which led to the result that the measure is not zero even when $(V_1, \cdots, V_m)$ is independent of $U$ due to the dependence within the group. With our construction, this issue is easily resolved.

In reality, that the upper bound of the dependence measure is not a constant or easily calculable may result in inconvenience in determining the true scale of dependence. As an alternative, we can calculate the dependence measures of each member of the group $(V_1, \cdots, V_m)$ on the other group $(U_1, \cdots, U_n)$ and average over all the values, as is the approach of Sanchez and Trutschnig (2014).

## 7. Conclusion

In this paper, we present a set of conditions, similar to the ones for bivariate nonsymmetric dependence measures in our previous paper (Li, 2015), that characterize the nonsymmetric measures of dependence between one continuous random variable and a group of continuous random variables. The measure takes value zero if and only if the one random variable is independent of the group of random variables and takes value one if and only if the one random variable is completely dependent on the group of random variables. Besides, the measure should be invariant under continuous bijective transformations of the random variables in the group. We explicitly construct new measures that satisfy the conditions. We also extend the $*$ product for bivariate copulas to multivariate case and use it to prove the DPI condition and self-equitability of the new measures. A further extension is made to measures of dependence of one group of random variables on another group of random variables.

A quick numeric test suggests that the dependence measures defined in this paper produces useful results. However, a more robust numeric scheme for handling higher order derivatives of copula is necessary for real applications. This should be the focus of future research.